\documentclass[12pt]{iopart}

\usepackage{iopams,bm}
\usepackage{psfrag,graphicx,epsfig,color}
\usepackage{graphicx,subfigure}
\usepackage[percent]{overpic}
\usepackage{float}
\usepackage{filecontents}

\usepackage{tikz}
\usetikzlibrary{arrows,intersections}
\usepackage{pgfplots}
\usepackage[all]{xy}

\begin{document}

\title{Lacunarity Transition}

\author{Bartomeu Cucurull$^{1}$, Marc Pradas$^{1}$ and Michael Wilkinson$^{1,2}$}

\address{$^{1}$ School of Mathematics and Statistics, The Open University,\\
              Walton Hall,  Milton Keynes, MK7 6AA, England.\\
             $^{2}$Chan Zuckerberg Biohub,499 Illinois Street,\\
             San Francisco, CA94158, USA\\}

\begin{abstract}

Experiments investigating particles floating on a randomly stirred fluid show regions of very low density, 
which are not well understood. We introduce a simplified model for understanding 
sparsely occupied regions of the phase space of non-autonomous, chaotic dynamical systems, based 
upon an extension of the skinny bakers' map. We show how the distribution of the sizes of voids 
in the phase space can be mapped to the statistics of the running maximum of a Wiener 
process. We find that the model exhibits a \emph{lacunarity transition},  which is characterised 
by regions of the phase space remaining empty as the number of trajectories is increased.

\end{abstract}

\section{Introduction}
\label{sec: 1}

\begin{figure}[h]
\centering
\includegraphics[width=0.65\textwidth]{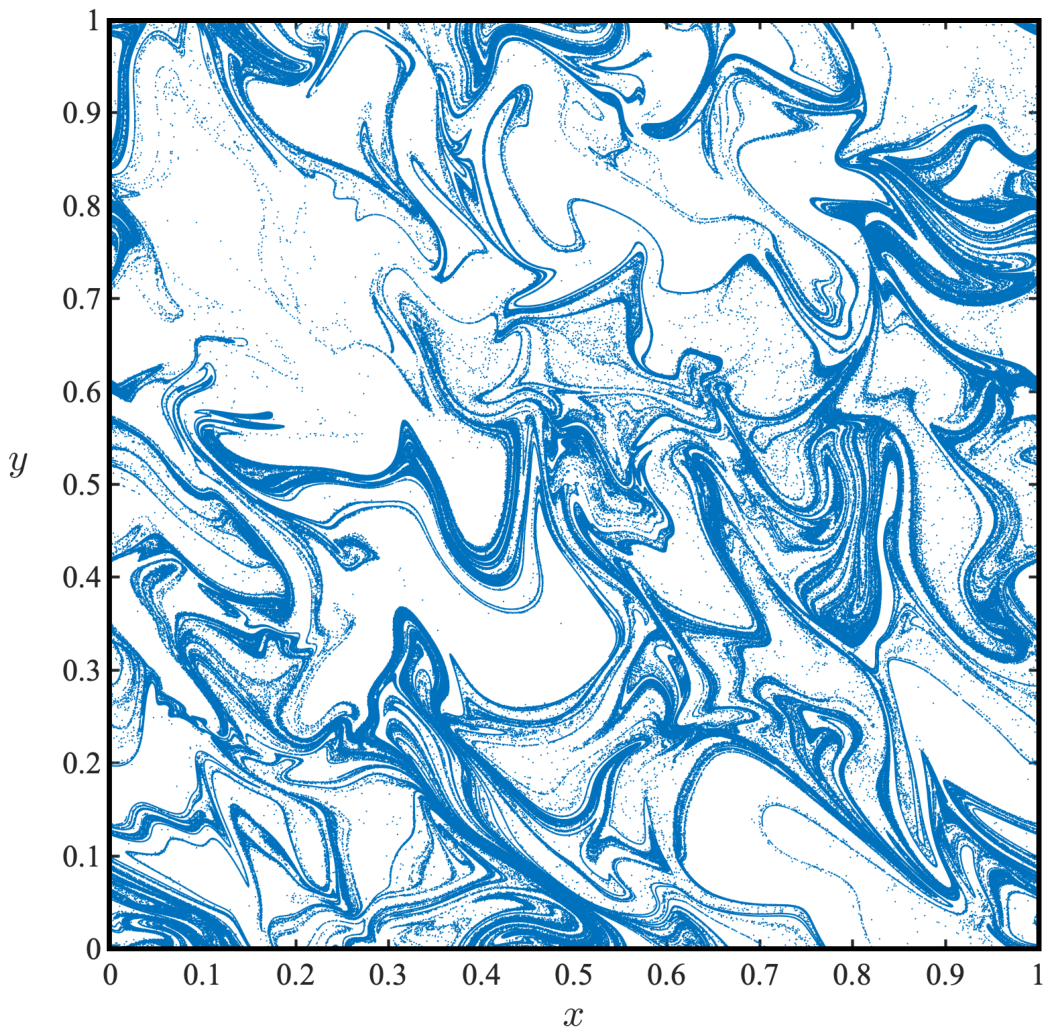}
\caption{
\label{fig: 1}
A model for particles floating on the surface of a randomly stirred fluid exhibits regions which 
are very sparsely occupied. This simulation represents the model discussed in \cite{Wil+18} 
(with compressibility parameter $\beta =0.5$ in the notation of that paper). 
We plot positions of $M=10^7$ trajectories, at a representative large time, which were initially a uniform 
random scatter. Note that there are substantial voids, which contain no trajectories.
}
\end{figure}

Very small particles floating on a chaotically stirred liquid \cite{Som+93,Lar+09} 
show regions where there is accumulation into regions of very high density, which are well 
described by fractal measures \cite{Som+93,Ott02}. These experiments also show regions of 
very low density, which were characterised in the paper by Larkin {\sl et al}, \cite{Lar+09}, 
but which are not yet well understood. Figure \ref{fig: 1} illustrates the \emph{lacunarity} 
of these chaotic attractors, by plotting $10^7$ trajectories of a dynamical system which 
mimics the motion of particles floating on the surface of a randomly stirred fluid
(the equations defining the model are the same as those considered in \cite{Bec+04,Wil+18},  
the model is precisely that considered in \cite{Wil+18} with compressibility parameter $\beta=1/2$). 
The concept of \emph{lacunarity}, characterising the tendency of some complex sets to 
have sparsely populated regions, was introduced by Benoit Mandelbrot in his classic 
book on fractals \cite{Man83}, but its influence has not been as far-reaching as the fractal dimension. 
This is perhaps because there is not a single agreed definition of how lacunarity should be 
quantified: see \cite{Plo+96} and \cite{Tol03} for a discussion of some definitions of lacunarity.

The fractal dimension concept has been extended to consider \lq multifractal' measures, 
which are considered to have different scaling exponent $\alpha$ in the vicinity of each point, 
and with the set of points with exponent $\alpha$ being a fractal with dimension $f(\alpha)$
\cite{Hal+86,Sal+17}. If this model is valid, the function $f(\alpha)$ is obtained by a Legendre
transform of the Renyi dimension, as discussed in \cite{Hal+86,Sal+17}. The extent to which our 
results are consistent with this model is considered in our concluding remarks, section \ref{sec: 6}. 
A recent paper \cite{Wil+19} considered sets arising as attractors of chaotic dynamical systems, 
and showed evidence that distribution of low densities has a power-law probability density function (PDF).
The exponent was termed the \emph{lacunarity exponent}. However, the model considered in that work 
was fundamentally different because its dynamics was many-to-one (as a result 
of \emph{folds} or \emph{caustics}), whereas here we consider a dynamical system which is \emph{invertible}. 
The theoretical arguments supporting the power-law described in \cite{Wil+19} are critically 
dependent upon the non-invertible nature of the systems which were considered there. 

This paper will introduce and analyse a simple model for invertible, non-autonomous, 
chaotic dynamical systems, such as the surface flow of a chaotically stirred fluid. 
The model is an extension of the skinny bakers' map, which is used 
as a minimal model for discussing fractality of chaotic attractors. Our model, which will be 
referred to as the \emph{strudel model}, differs from the skinny bakers' map in two respects.
Firstly, unlike the skinny bakers' map, it is invertible: there are no 
inaccessible regions of the phase space. Secondly, 
the discontinuities are introduced at random positions. Introducing this 
random element has two advantages. Firstly, it makes the phase space statistically homogeneous. 
Secondly, the randomness facilitates our analysis of the system by enabling the use of statistical methods. 

Here we describe sparse regions by considering the distribution of $M\gg 1$ trajectories, and 
considering the statistics of the size $\epsilon$ of the 
trajectory-free void surrounding an arbitrarily chosen point. 
We characterise the distribution of $\epsilon$ by determining how the expectation value
of its logarithm, $\langle \ln \epsilon \rangle$ varies as a function of $\ln M$. 
We show that the distribution of $\ln \epsilon$ may be mapped to determining the running 
maximum of a Wiener process with drift.

At very large values of $M$ there is a linear dependence: $\langle \ln \epsilon\rangle\sim -\gamma \ln M$, 
for some exponent $\gamma$, which depends upon the 
parameters of the model.  We find that the value of $\gamma$ is equal to zero for some regions. 
When $\gamma$ becomes equal to zero, the voids in the distribution 
of trajectories are not filled when we add more trajectories, whereas the voids are filled 
by adding more trajectories when $\gamma>0$. We say that the edge of the region where $\gamma=0$ 
marks a phase transition, which we term the \emph{lacunarity transition}.

Section \ref{sec: 2} will introduce the strudel model, and describe its backward iteration as well as
forward iteration. Section \ref{sec: 3} discusses a succession of models for distribution of the the void 
size $\epsilon$, and section \ref{sec: 4} discusses the lacunarity transition, where the distribution 
of $\epsilon$ changes abruptly in the limit as the number of trajectories $M$ is increased. Section 
\ref{sec: 5} discusses our numerical results, which show good agreement with the theory of section 
\ref{sec: 3}, despite the quite brutal coarse-graining approximations which are used. Section \ref{sec: 6}
contains some concluding remarks on the relation to earlier work and prospects for extension of the theory 
to more physically realistic models.

\section{Strudel model}
\label{sec: 2}

\subsection{Definition of model}
\label{eq: 2.1}

The skinny bakers' map \cite{Ale+84} is a piecewise linear map, which mimics the 
stretch-and-fold action of a typical chaotic system. The unit square is stretched to twice its length in the $x$-direction, 
whilst being contracted by more than a factor of two in the $y$-direction. The stretched region is then 
cut into two halves which are placed in the upper and lower halves of the unit square. The unit square 
is, therefore, mapped into two rectangles, both of dimension $1\times \beta/2$, where $\beta\in[0,1]$. The resulting 
attractor is the Cartesian product of the unit interval and a fractal Cantor set. The fractal dimension 
of the attractor is $d=1+\ln 2/(\ln 2-\ln \beta)$. 

Our model is a variant of this skinny baker map, which we shall 
refer to as the \emph{strudel model}. It is an invertible two-dimensional random dynamical 
system, which is designed to have regions of very low density, and to be simple 
enough to facilitate making analytical approximations to the distribution of sizes of empty regions.
It also has the advantage that, by virtue of being a map rather than a flow, it is suited
to efficient numerical work. The operation of the map is illustrated schematically by figure \ref{fig: 2}.
The map depends upon two parameters, $p\in[0,1]$ and $\beta\in[0,1]$. 
It acts on a point $(x,y)$ in the unit square, as described by equations (\ref{eq: 2.1.1}) and (\ref{eq: 2.1.2}) below.
In the first step, a unit square is subjected to a continuous, piecewise 
linear, transformation of the $y$ component. The square is then stretched 
by a factor of $2$ in the $x$-coordinate, and contracted by a factor of $2$
in the $y$-direction. The $2\times 1/2$ rectangle is then cut into two halves, which are 
moved back into the unit square. 

\begin{figure}[t!]
\centering
\bigskip 

\begin{tikzpicture}
\draw [black] (0.0,0.0) rectangle ++(1.5,1.5);
\draw [fill=lightgray,lightgray] (0.0,0.1) rectangle ++(1.5,1.3);

\draw[->] (1.7,0.75) -- (2.3,0.75);

\draw [black] (2.5,0.0) rectangle ++(1.5,1.5);
\draw [fill=gray,gray] (2.5,0.1) rectangle ++(1.5,0.8);

\draw[->] (4.2,0.75) -- (4.8,0.375);

\draw [black] (5.0,0.0) rectangle ++(3.0,0.75);
\draw [fill=gray,gray] (5.0,0.05) rectangle ++(3.0,0.4);

\draw[->] (8.2,0.375) -- (8.8,0.75);

\draw [black] (9.0,0.0) rectangle ++(1.5,1.5);
\draw [fill=gray,gray] (9.0,0.05) rectangle ++(1.5,0.4);
\draw [fill=gray,gray] (9.0,0.80) rectangle ++(1.5,0.4);

\node at (0.75,0.75) {$1-p$};
\node at (3.25,0.5) {$\beta$};
\node [left] at (0.0,0.1) {$\phi_n$};

 \end{tikzpicture}
\caption{
Illustrating the action of the strudel map. 
At the first step, there is a continuous, piecewise linear transformation of the $y$-coordinate of the 
unit square, which maps a region of length $1-p$ to length $\beta$. The lower edge
of this region is at a random position, $\phi_n$. This region is then stretched along the $x$-axis, to occupy 
a $2\times 1/2$ rectangle. This rectangle is cut and the two halves are stacked back into the unit square.
} 
\label{fig: 2}
\end{figure}

To describe the transformation of the $y$-coordinate, we define a periodic function, $F(x)=F(x+1)$ by 
specifying its values on $[0,1]$ as follows:
\begin{equation}
\label{eq: 2.1.1}
F(x)=\left\{
\begin{array}{cc}
\frac{\beta}{1-p}x& x\in [0,1-p]\cr
\beta+\frac{1-\beta}{p}(x+p-1)& x\in [1-p,1]
\end{array}
\right.
\ .
\end{equation}
Also let $n$ be the index of the iteration and let $\phi_n$ be a random number, uniform on $[0,1]$, 
chosen independently at each iteration. Then we define the \emph{strudel map} as follows:
\begin{eqnarray}
\label{eq: 2.1.2}
x_{n+1}&=&2x_n\,{\rm mod}\, 1
\nonumber \\
y_{n+1}&=&\frac{1}{2}\left[F(y_n-\phi_n)+{\rm int}(2x_n)\right]
\ .
\end{eqnarray}
If we set $\phi_n=0$ and $p=0$, this is the skinny baker map \cite{Ale+84}, which 
has empty regions which occupy a fraction $1-\beta^N$ of the phase space 
after $N$ iterations. When $p>0$, there are no inaccessible regions, but as $p\to 0$ 
the density of some regions may be very small.

\begin{figure}[t]
\centering
\includegraphics[width=0.5\textwidth]{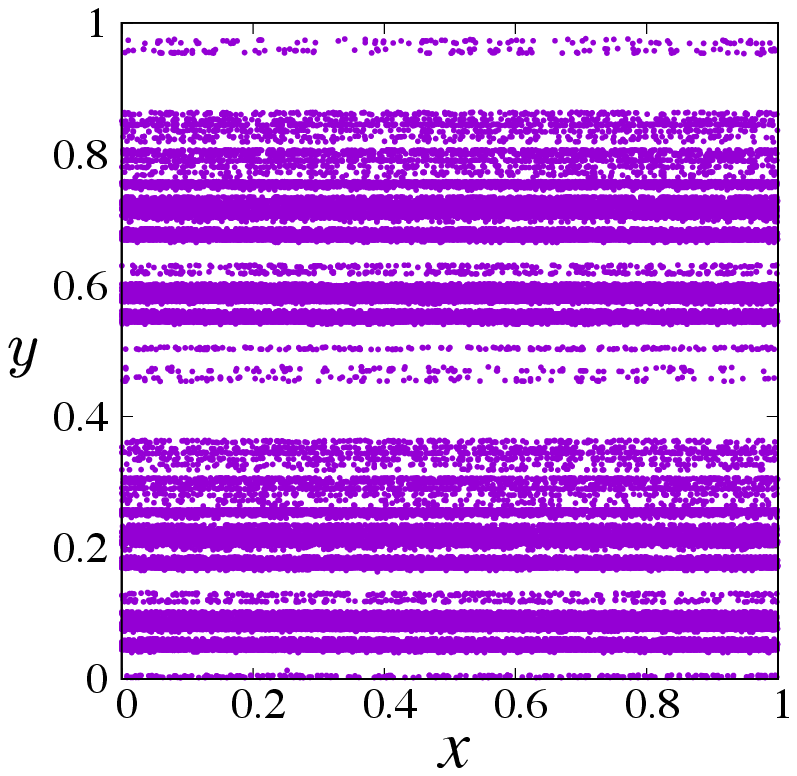}
\caption{
\label{fig: 3}
Distribution of trajectories for a realisation of the strudel map. 
The parameter values are $p=0.2$, $\beta=0.4$. We mapped $M=10^5$ 
randomly scattered initial conditions for $N=15$ iterations. Note that these 
are some substantial gaps in the distribution of the $y$-coordinate.
}
\end{figure}

For completeness, we give expressions for the Lyapunov exponents of this model
and its fractal dimensions. Small separations in the $x$-coordinate are doubled upon each iteration. 
If we also define
\begin{equation}
\label{eq: 2.1.3}
\xi_1=\ln\left(\frac{1-\beta}{2p}\right)
\ ,\ \ \ 
\xi_2=\ln\left(\frac{\beta}{2(1-p)}\right)
\end{equation}
then the logarithm of the small separation in the $y$-coordinate is incremented by either $\xi_1$ or $\xi_2$ 
with probability $p$ or $1-p$ respectively. The Lyapunov exponents are therefore
\begin{equation}
\label{eq: 2.1.4}
\lambda_1=\ln 2
\ ,\ \ \ 
\lambda_2=p\xi_1+(1-p)\xi_2
\ .
\end{equation}
Of the Renyi dimensions, two are easily determined. If $p>0$, the box counting dimension is $d_0=2$
because there are no inaccessible points. And the information dimension as estimated 
by the Kaplan-Yorke formula  \cite{Kap+79} is 
\begin{equation}
\label{eq: 2.1.5}
d_1=1+\frac{\lambda_1}{|\lambda_2|}
=1+\frac{\ln 2}{\bigg\vert p\ln\left(\frac{1-\beta}{2p}\right)+(1-p)\ln\left(\frac{\beta}{2(1-p)}\right)\bigg\vert}
\ .
\end{equation}
The distribution of points generated by this map is a random scatter in the $x$-direction, 
but highly inhomogeneous in the $y$-coordinate. An example is shown in figure \ref{fig: 3}.
The striated texture of this image resembles the fine structure of the foliations shown in figure \ref{fig: 1}. 

\subsection{Pre-images}
\label{sec: 2.2}

Consider the distribution obtained from $M\gg 1$ trajectories, which are initially 
uniformly scattered on the unit square, after $N\gg 1$ iterations. These end up 
uniformly scattered in the $x$-coordinate, but the values of the $y$-coordinate are 
highly inhomogeneous, as illustrated in figure \ref{fig: 3}. Let us sort the $y$-coordinates 
of the trajectories into ascending order. If we then pick a point at random, it can be placed 
inside a rectangular void, of dimensions $1\times \epsilon$, with trajectories on the upper 
and lower edges. The values of the gap size $\epsilon$ in the $y$-coordinate are random variables. We can 
characterise the lacunarity of the distribution of trajectories by determining the PDF of $\epsilon$, 
or by determining its statistics.  

To understand the statistics of these void regions, notice that all of the pre-images of a void 
are also empty regions. If we follow the evolution backwards by $N$ steps to the initial
configuration, all of the pre-images are also empty. After $n$ steps backwards, the area 
of the pre-image of the $1\times \epsilon$ rectangle is denoted by $A_n$, and the area
of the initial empty region is $A_N$. Because the initial distribution is a random scatter 
of $M$ points in the unit square, the probability of an area $A$ in the initial configuration 
being empty is $P(A)=\exp(-MA)$, so that the probability of the area of the $N$ step 
pre-image being a large multiple of $1/M$ is very small. This implies that 
the large gaps in the $y$-coordinate arise as a consequence of having small areas of the 
pre-image. 

Let us consider the sequence of pre-images of a rectangular region of size 
$1\times \epsilon$ which has its lower edge at $y$, after $N$ steps backwards. 
To facilitate the discussion, we first obtain an expression for the pre-image of a point. 
The forward map is defined by equations (\ref{eq: 2.1.2}), with $F(x)$ defined by (\ref{eq: 2.1.1}). 
We define a function $G$ that is the inverse of $F$:
\begin{equation}
\label{eq: 2.2.1}
G(F(x))=x
\end{equation}
that is
\begin{equation}
\label{eq: 2.2.2}
G(x)=\left\{
\begin{array}{cc}
\frac{1-p}{\beta}x& x\in [0,\beta]\cr
1-p+\frac{p}{1-\beta}(x-\beta)& x\in [\beta,1]
\end{array}
\right.
\ .
\end{equation}
We extend the definition of $G(x)$ to the whole real line as a periodic function with unit period.
Noting that equation (\ref{eq: 2.1.2}) implies that ${\rm int}(2x_n)={\rm int}(2y_{n+1})$,  
we have $2y_{n+1}-{\rm int}(2y_{n+1})=F(y_n-\phi_n)$, so acting on this relation with $G$ we obtain
\begin{equation}
\label{eq: 2.2.3}
y_n=\phi_n+G(2y_{n+1})-{\rm int}(2y_{n+1})
\end{equation}
where we use the fact that $=G({\rm int}(2y_{n+1}))={\rm int}(2y_{n+1})$.  
This has the nice feature that it is independent of the $x_n$ coordinate. 
Also, because the function $G$ has been constructed to be periodic with 
unit period, we can simplify further by applying the following backward iteration:
\begin{equation}
\label{eq: 2.2.4}
\tilde y_n=\phi_n+G(2\tilde y_{n+1})
\end{equation}
and recover the value of $y_n$ by subtracting the integer part.
The backward iteration of the $x_n$ coordinate is a little more complicated: the pre-image 
of any vertical line which crosses the horizontal line $y=1/2$ consists of two segments, 
with horizontal separation equal to one half. However, the pre-images of a rectangle 
always reduce in width by a factor of two with each iteration.
 
Consider the backward iteration of an $1\times \epsilon$ rectangular region, where the 
lower and upper edges are two successive values of the $y$-coordinate after $N$ iterations, 
differing by $\epsilon$, with the lower edge at $y_n$.  After $N$ backward steps, this 
maps to a set of rectangular regions, each one of which has width $\Delta x=2^{-N}$.  
The sum of the vertical extent of each fragment is $\Delta \tilde y$, which is iterated according to 
\begin{equation}
\label{eq: 2.2.5}
\Delta \tilde y_{n}=G(2y_{n+1}+2\Delta \tilde y_{n+1})-G(2y_{n+1})
\end{equation}
starting with $\Delta \tilde y_N=\epsilon$. The pre-image of the $1\times \epsilon$ rectangle 
is a set of rectangular regions of total area 
\begin{equation}
\label{eq: 2.2.6}
A_{n}=2^{-N}\Delta \tilde y_n
\ .
\end{equation}
If $\epsilon \ll 1$, the iteration of (\ref{eq: 2.2.5}) can be 
approximated by linearisation, so that after $N$ steps of backward iteration 
the total vertical extent of the pre-image area is  
\begin{equation}
\label{eq: 2.2.7}
\Delta \tilde y_N\sim 2^N\, \epsilon\, \prod_{i=1}^N G'(2y_n)
\ .
\end{equation}
Thus $\Delta \tilde y_n$ typically grow under iteration, and the approximation (\ref{eq: 2.2.7})
ceases to be valid when $\Delta \tilde y$ is of order one. When 
$\Delta \tilde y_n\gg 1$, we use the fact that 
\begin{equation}
\label{eq: 2.2.8}
\int_0^1 {\rm d}x\ G'(x)=1  
\end{equation}
and conclude that $\Delta \tilde y_n 2^{-N}$  becomes independent of $N$ for sufficiently 
large $N$. Given that $\Delta x_N=2^{-N}$, this implies that the area $A_N$ of the pre-image set 
approaches a constant as $N\to \infty$.

We note that the Lyapunov exponents for the backward propagation are different from the forward
Lyapunov exponents. Defining $\bar \xi=\ln 2G'$, we see that $\bar \xi$ takes two 
possible values, which occur randomly and independently in the sequence of $y_n$ values:
\begin{eqnarray}
\label{eq: 2.2.9}
\bar \xi_1=\ln \left(\frac{2(1-p)}{\beta}\right)&\quad&{\rm probability}\ p_1=\beta
\nonumber \\
\bar \xi_2=\ln \left(\frac{2p}{1-\beta}\right)&\quad&{\rm probability}\ p_2=1-\beta
\ .
\end{eqnarray}
The Lyapunov exponents of the backward iterated map are then
\begin{equation}
\label{eq: 2.2.10}
\bar \lambda_1=\beta \bar \xi_1+(1-\beta)\bar \xi_2
\ ,\ \ \ 
\bar \lambda_2=-\ln 2
\ .
\end{equation}

\section{Model for distribution of void sizes}
\label{sec: 3}

\subsection{Representation in logarithmic variables}
\label{sec: 3.1}

Consider the pre-image of a rectangular region of size $1\times \epsilon$ after 
$n$ backwards iterations. It is mapped to a set of rectangular regions of total area 
$A_n=\Delta x\times \Delta \tilde y$. While $\Delta \tilde y_n\ll 1$ its evolution is well approximated by 
\begin{equation}
\label{eq: 3.1.1}
A_n\sim \epsilon \prod_{j=1}^n G'(2y_j)
\end{equation}
where $G'(2y_j)$ takes one of two values, $(1-p)/\beta$ or $p/(1-\beta)$, with probabilities 
$p_1=\beta$ or $p_2=1-\beta$, respectively. After $\Delta \tilde y$ has grown to be 
of order unity, the area $A_n=\Delta x_n\Delta \tilde y_n$ of the pre-image set stabilises, at a value 
denoted by $\tilde A$. The size of the open interval, $\epsilon$, is determined by the condition 
that $\Delta \tilde y_n$ never exceeds unity, while the area of the pre-image reduces to $1/M$ or less. 

It is convenient to use logarithmic variables:
\begin{equation}
\label{eq: 3.1.2}
X_1=\ln \Delta x
\ ,\ \ \ 
X_2=\ln \Delta \tilde y
\ .
\end{equation}
The backwards evolution of $X_1$ is trivial, and the evolution of $X_2$ follows from equation
(\ref{eq: 2.2.7}): after $N$ backwards steps we have
\begin{equation}
\label{eq: 3.1.3}
X_1=-N \ln 2 
\ ,\ \ \ 
X_2=\ln \epsilon+N\sum_{j=1}^N \bar \xi_j
\end{equation}
where the $\bar \xi_j$ take one of two values as specified by equation (\ref{eq: 2.2.9}).

Note that the condition $\Delta \tilde y\le 1$ corresponds to the constraint
$X_2\le 0$. In terms of the logarithmic variables, the condition that $\tilde A\le 1/M$ is 
\begin{equation}
\label{eq: 3.1.4}
X_1+X_2\le -\ln M\ ,
\end{equation}
and the dynamical process describing the evolution 
of the pre-image is therefore a random 
walk in $X_2$, as a function of $X_1$. The initial condition is $(X_1,X_2)=(0,\ln\epsilon)$.
The point moves to the left in $(X_1,X_2)$ space by $\ln 2$ at each step. The motion proceeds until 
(\ref{eq: 3.1.4}) is satisfied, and we choose the largest value of $\epsilon$ so that 
$X_2$ never exceeds zero. When $\Delta \tilde y=1$, the area is $A_N=\Delta x\Delta \tilde y=2^{-N}$, 
so that the number of backward iterations is 
\begin{equation}
\label{eq: 3.1.5}
N=\frac{\ln M}{\ln 2}
\end{equation}
(which achieves $\tilde A=1/M$) or greater (which results in a smaller pre-image).  
The trajectory in $(X_1,X_2)$ space is illustrated in figure \ref{fig: 4}.

\begin{figure}[t!]
\centering 
\resizebox{0.7\textwidth}{!}{
\begin{tikzpicture}
\draw [thick, gray, ->] (0.0,8.0) -- (9.0,8.0)  
        node [right, black] {$X_1=\ln(\Delta x)$};  
\draw [thick, gray, ->] (8.0,0.0) -- (8.0,9.0)  
        node [right, black] {$X_2=\ln(\Delta y)$}; 
\draw [thick, gray] (0.75,8.25) -- (8.5,0.5)  
        node [right, black] {$X_1+X_2=-\ln(M)$}; 
\draw [thick,black](1.0,7.4) -- (1.5,7.7);
\draw [thick,black](1.5,7.7) -- (2.0,8.0);
\draw [thick,black](2.0,8.0) -- (2.5,7.5);
\draw [thick,black](2.5,7.5) -- (3.0,7.8);
\draw [thick,black](3.0,7.8) -- (3.5,7.3);
\draw [thick,black](3.5,7.3) -- (4.0,6.8);
\draw [thick,black](4.0,6.8) -- (4.5,7.1);
\draw [thick,black](4.5,7.1) -- (5.0,6.6);
\draw [thick,black](5.0,6.6) -- (5.5,6.9);
\draw [thick,black](5.5,6.9) -- (6.0,6.4);
\draw [thick,black](6.0,6.4) -- (6.5,5.9);
\draw [thick,black](6.5,5.9) -- (7.0,5.4);
\draw [thick,black](7.0,5.4) -- (7.5,5.7);
\draw [thick,black](7.5,5.7) -- (8.0,5.2);

\node at (1.0,8.0)[circle,fill,inner sep=2pt]{};
\node at (8.0,5.2)[circle,fill,inner sep=2pt]{}; 
\node [right] at (8.0,5.2) {$(0,\ln \epsilon$};  
\node [above] at (1.0,8.0) {$(-N\ln(2),0)$};  
 
\end{tikzpicture}
}
\caption{
Schematic illustration of the dynamics determining void size, expressed 
in logarithmic coordinates, equation (\ref{eq: 3.1.2}). The trajectory starts from 
$(0,\ln \epsilon)$ and makes a biased random walk, until it exits the triangular region
$X_1+X_2<-\ln M$. The value of $\epsilon$ is chosen so that the trajectory never enters
the region $X_2>0$.
}
\label{fig: 4}
\end{figure}

\subsection{Modelling by Wiener process}
\label{sec: 3.2}

Next we make a further approximation, which enables 
us to approximate the statistics of the void sizes by simple analytic formulae.
The motion of $X_2$ as a function 
of $X_1$ defined by equation (\ref{eq: 3.1.3}) is a biased random walk. It can be modelled as a 
Wiener process, $x(t)$ where $t\equiv -X_1$ and $x\equiv X_2$. This Wiener process has a 
drift velocity $v$ and a diffusion coefficient $D$. The mean and variance of the change in 
$x$ over one timestep, $\Delta t=\ln 2$, are $v\Delta t=p_1\bar \xi_1+p_2\bar \xi_2$, and 
$2D\Delta t=p_1\bar \xi_1^2+p_2\bar \xi_2^2-v^2\Delta t^2$, so that
\begin{equation}
\label{eq: 3.2.1}
v=\frac{p_1\bar \xi_1+p_2\bar\xi_2}{\ln 2}=\frac{\bar \lambda_1}{\ln 2}
\end{equation}
and 
\begin{equation}
\label{eq: 3.2.2}
D=\frac{1}{2\ln 2}\left[p_1\bar\xi_1^2+p_2\bar\xi_2^2-(p_1\bar\xi_1+p_2\bar \xi_2)^2\right]
\ .
\end{equation}
For each realisation of the Wiener process, we must determine the largest value of $x_0=\ln \epsilon<0$
such that if $x(t)$ starts at $x_0$, it remains negative for all times $t$ up to 
\begin{equation}
\label{eq: 3.2.3}
T=N\ln 2=\ln M
\ .
\end{equation}
Alternatively, $-\ln \epsilon$ is the maximum value of a Wiener process $x(t)$ in the time interval $t\in [0,T]$. 
This is illustrated schematically in figure \ref{fig: 5}. 

\begin{figure}[t!]
\centering 
\resizebox{0.65\textwidth}{!}{
\input{fig5.tex}
}
\caption{
\label{fig: 5}
Schematic illustration of the dynamics determining void size, expressed 
in logarithmic coordinates, where the biased random walk is approximated 
by a Wiener process. We require the statistics of the running maximum of 
the Wiener process $x(t)$, up to time $T$. 
}
\end{figure}

\subsection{Estimate for mean value}
\label{sec: 3.3}

Now let us estimate the mean value of $x_0=\ln \epsilon$, using the 
Wiener process model. If the diffusion coefficient were $D=0$, and $v>0$, and we were to 
release a particle at $x_0=-vT$, then it would reach $x=0$ when $t=T$.  In this deterministic case we would have 
$\langle x_0\rangle=-vT$. On the other hand, if $v=0$ we would expect that the maximum 
displacement would be of order $\sqrt{DT}$. If diffusion is significant, but $v\ne 0$, we might, 
therefore, anticipate that
\begin{equation}
\label{eq: 3.3.1}
\langle x_0\rangle=-\sqrt{2DT}\  F(Y)
\end{equation}
where $F(Y)$ is a function of a dimensionless variable
\begin{equation}
\label{eq: 3.3.2}
Y=v\sqrt{\frac{T}{2D}}
\end{equation}
and where $F(Y)\sim Y$ as $Y\to \infty$.
In the Appendix we show that the function $F(Y)$ is
\begin{equation}
\label{eq: 3.3.3}
F(Y)=\Phi'(Y)+\Phi(Y)\frac{1+Y^{2}}{Y}-\frac{1}{2Y}
\end{equation}
where
\begin{equation}
\label{eq: 3.3.4}
\Phi(x)=\frac{1}{\sqrt{2\pi}}\int_{-\infty}^x{\rm d}y\ \exp(-y^2/2)
\end{equation}
is the cumulative distribution function of a Gaussian with unit variance. The limiting behaviours of $F(Y)$ are 
\begin{equation}
\label{eq: 3.3.5}
F(Y)\sim \left\{
\begin{array}{cc}
Y&Y\gg 1\cr
\sqrt{\frac{2}{\pi}}&Y=0\cr
\frac{1}{2|Y|}&-Y\gg 1
\end{array}
\right.
\ .
\end{equation}

\section{Lacunarity transition}
\label{sec: 4}

We have proposed a theory for the statistic $\langle \ln \epsilon \rangle$, 
where $\epsilon$ characterises the size of a void region. In the limit as the number of trajectories 
$M$ approaches infinity, the dimensionless variable $Y$ defined by (\ref{eq: 3.3.2}) is large, and 
(according to equations (\ref{eq: 3.3.1}) and (\ref{eq: 3.3.5}) the theory predicts that  
\begin{equation}
\label{eq: 4.1}
\langle \ln \epsilon \rangle \sim -v \ln M
\end{equation}
where $v$ is given by equation (\ref{eq: 3.2.1}), provided that $v>0$. 
This is consistent with the typical size of $\epsilon$ having a 
power-law dependence: 
\begin{equation}
\label{eq: 4.2}
\epsilon \sim M^{-\gamma}
\end{equation}
where the exponent is $\gamma=v$.
If $\gamma<1$ this indicates that the voids are larger than would be expected 
for a random scatter of points, for which the separation of the ordered $y$-coordinates 
would be $\epsilon\sim 1/M$.

In the case where $v<0$, however, $Y\to -\infty$ as $M\to \infty$, and the theory 
predicts that $\langle \ln\epsilon\rangle $ becomes independent of $M$ as $M\to \infty$,
so that $\gamma=0$ in regions where $v<0$. 
There is indeed a region in the parameter space of our model where $v<0$. In this case 
$\langle \ln \epsilon\rangle\sim D/ v $, which is independent of $M$. This implies 
that when $v<0$, there are voids in the distribution of trajectories which are not filled 
as we increase their number. The locus where $v=0$ in the parameter space of the 
model $\{p,\beta\}\in [0,1]^2$ represents a phase transition, between a phase space which 
fills every region as $M\to \infty$ when $v>0$, to one which has persistent voids when $v<0$.

\begin{figure}[t]
\centering
\includegraphics[width=0.98\textwidth]{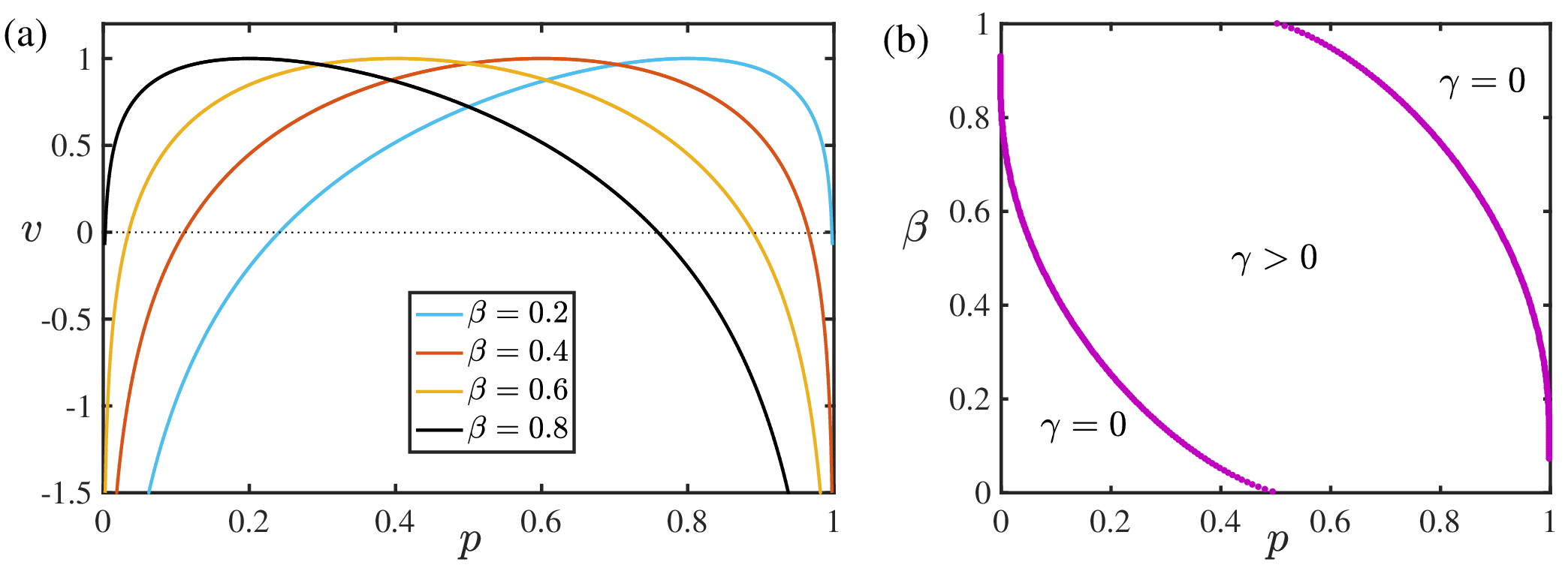}
\caption{
(a) Plots of $v$ as a function of $p$ for $\beta=0.2,0.4,0.6,0.8$. The dotted line 
indicates $v=0$ for reference. (b) Plot of the $(p,\beta)$ parameter space of the model. 
The purple lines correspond to the locus of the phase transition, with $\gamma>0$ in 
the region between the lines and $\gamma = 0$ everywhere else.}
\label{fig: 6}
\end{figure}

The value of $v$ as a function of $p$ for different choices of $\beta$ is 
shown in figure \ref{fig: 6}(a). The line of the phase transition  in the $(p,\beta)$ parameter 
space is illustrated  in figure \ref{fig: 6}(b). 

\section{Numerical simulations}
\label{sec: 5}

\begin{figure}[t]
\centering
\includegraphics[width=0.7\textwidth]{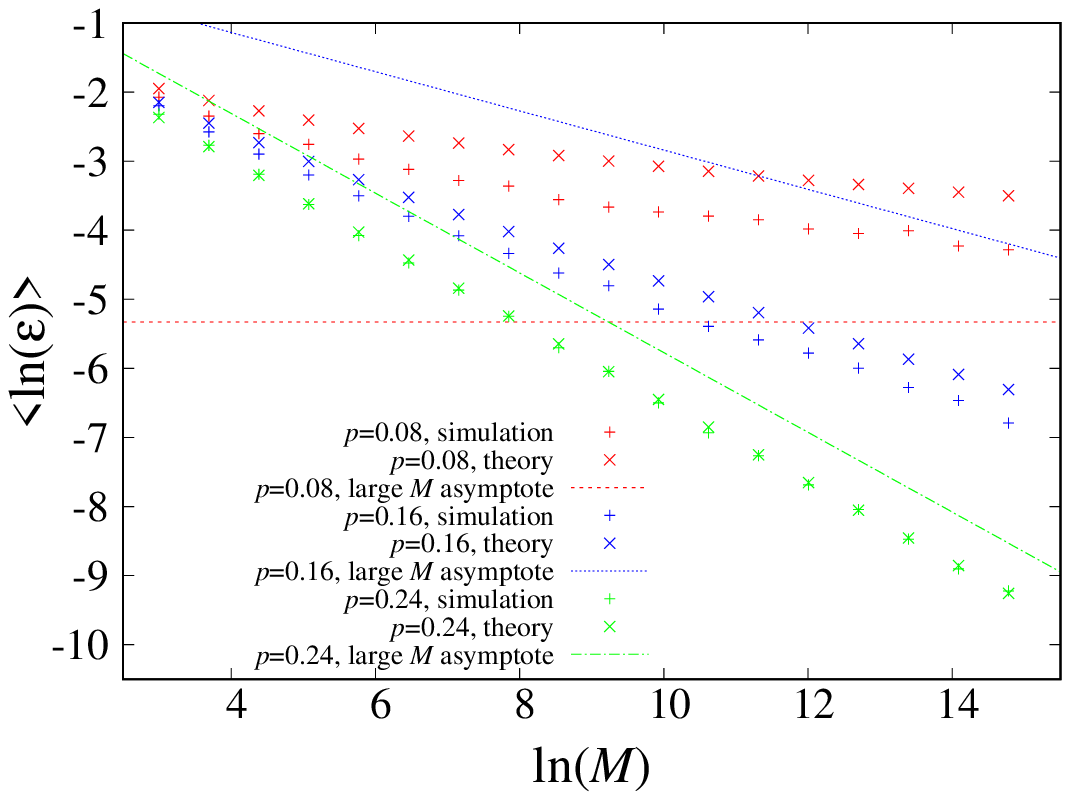}
\caption{
\label{fig: 7}
Plots of $-\langle \ln \epsilon\rangle $ as a function of $T=\ln M$, for $\beta=0.4$, with three 
different values of $p$. These data are compared with the theory, equations (\ref{eq: 3.3.1}) and (\ref{eq: 3.3.3}). 
The dotted lines are the asymptotes of the theory for very large values of $M$, showing that 
even $10^6$ trajectories are not sufficient to explore the $M\to \infty$ limit. 
}
\end{figure}

We evaluated the values of $-\langle \ln \epsilon\rangle$,  after $N=100$ iterations of the map. 
We averaged $N_{\rm r}=500$ realisations of the map, and in each case we evaluated the set of 
void sizes $\epsilon$ using $K=250$ evenly spaced initial points. 
The resulting expectation values of $\ln\epsilon$ are compared 
with the theoretical prediction, equations (\ref{eq: 3.3.1})
and (\ref{eq: 3.3.3}) in figure \ref{fig: 7} for three different points in the parameter space of the 
model. The agreement between the simulations and equation (\ref{eq: 3.3.1}) is excellent in one case 
($\beta=0.4$ and $p=0.24$), but the other two cases show a small offset between the simulation and 
this theoretical prediction, which is approximately independent of $M$.

The values of the drift velocity and diffusion 
coefficient for the three cases illustrated in figure \ref{fig: 7} are:
\begin{eqnarray} 
{\rm for\ }p=0.08,\ \beta=0.4&\quad \rightarrow \quad&v=-0.2634\ldots,\ D=1.4040\ldots
\nonumber \\
{\rm for\ }p=0.16,\ \beta=0.4&\quad \rightarrow \quad&v=0.2840\ldots,\ \ \ D=0.7373\ldots
\nonumber \\
{\rm for\ }p=0.24,\ \beta=0.4&\quad \rightarrow \quad&v=0.5772\ldots,\ \ \ D=0.4203\ldots
\nonumber \\
\end{eqnarray}
When $v>0$, the asymptotic behaviour as $M\to \infty$ is $\langle \ln \epsilon\rangle\sim -v\ln M$, 
whereas if $v<0$,  $\langle \ln \epsilon\rangle\sim D/v$. 
The data points in figure \ref{fig: 7} all appear to be well approximated by a straight line when $M$ is large. 
However, this figure also shows the $M\to \infty$ asymptotic behaviour as lines, and even for 
the largest values of $M$ (up to $10\times 2^{18}>10^6$), the 
theoretical expression is still far from these asymptotes.
We conclude that the true asymptotic behaviour is not accessible even for the very large values of $M$ 
which are explored in figure \ref{fig: 7}.

\begin{figure}[t]
\centering
\includegraphics[width=0.7\textwidth]{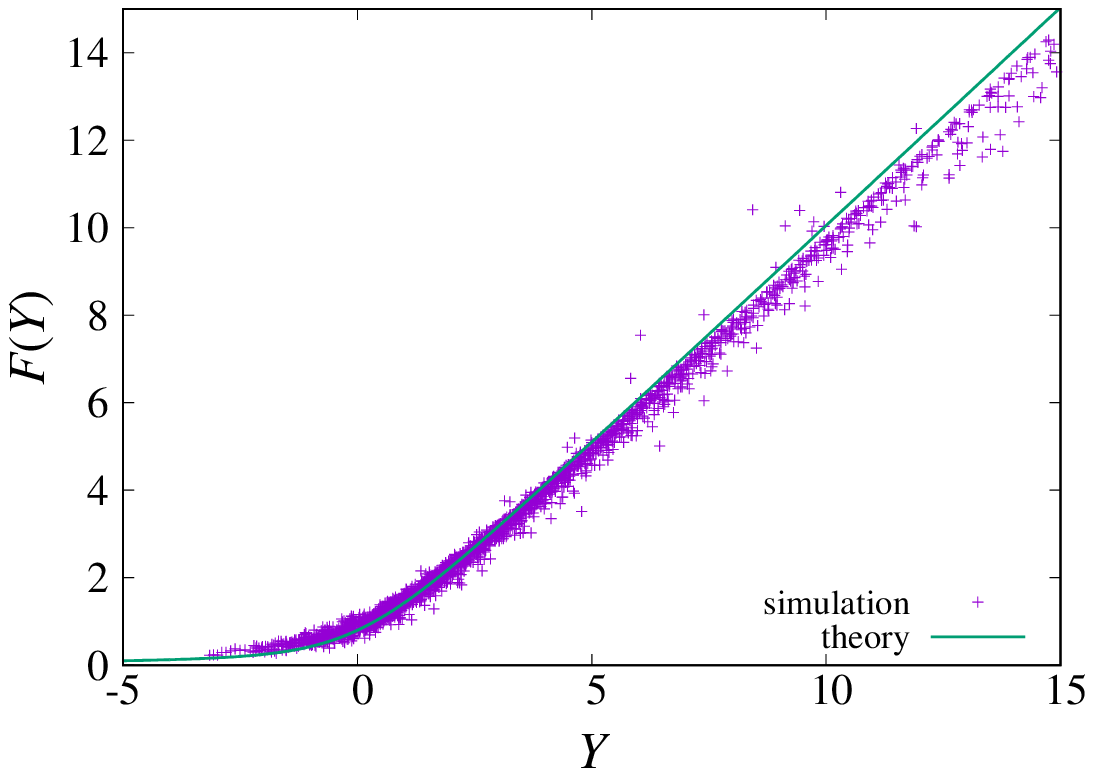}
\caption{
\label{fig: 8}
Scatterplot of $Z=-\langle \ln \epsilon\rangle/\sqrt{2DT}$ as a function of $Y=v\sqrt{T/2D}$, for more than $10^3$ different combinations of values of $p$, $\beta$ and $T=\ln M$.
These data are compared with the function $F(Y)$ (solid line), defined by equation (\ref{eq: 3.3.3}).  
}
\end{figure}

We also evaluated $\langle \ln \epsilon\rangle $ for $18$ different values of $M$, namely 
$M=4,8,\ldots,2^{19}$, for all values of $p$ and $\beta$ forming a lattice in the parameter 
space. (The lattice spacing was $0.075$, with $p$ taking values from $0.075$ to $0.9$ and $\beta$ 
from $0.15$ to $0.9$, making $132$ different points in the parameter space). For each of these 
$132\times 18$ data points we determined $v$ and $D$ from the values of $p$ and $\beta$ and $T=\ln M$. 
We then computed $Y=v\sqrt{T/2D}$ and $Z=-\langle \ln \epsilon\rangle/\sqrt{2DT}$. 
Figure \ref{fig: 8} is a scatterplot  of $Z$ against $Y$, compared with the function $F(Y)$, 
given by equation (\ref{eq: 3.3.3}). 
There is a good scaling collapse of the scatterplot onto a single line, and this 
line is in good agreement with the function $F(Y)$.

\section{Concluding remarks}
\label{sec: 6}

Data from both physical and numerical experiments on non-autonomous chaotic systems indicate that  there can be very sparsely occupied regions of phase space. These have previously been investigated 
for the case of systems which have folds or caustics \cite{Wil+19}, but in the case of systems with 
invertible dynamics there is very little previous work.

In this paper we considered a simple model, which is susceptible to analysis by mapping the 
problem to that of determining the running maximum of a biased diffusion process. 
The model system which we consider has uniform distribution of trajectories in its $x$-coordinate, 
but a highly non-uniform distribution of the $y$-coordinate, as illustrated in figure \ref{fig: 3}. We considered $M$ trajectories with uniformly scattered initial points, after $N$ iterations of the map.  A randomly chosen point $(x,y)$ can be positioned in a rectangle of dimensions $1\times \epsilon$, which contains none of the trajectories in its interior, but which does have one trajectory on its upper and lower edges. The statistics of the gap size, $\epsilon$, provide a means to describe figure \ref{fig: 3}.  We developed a theory for $\langle \ln \epsilon\rangle$, predicting that $\langle \ln \epsilon \rangle\sim -\gamma\ln M$
as $\ln M\to\infty$, where $\gamma$ is a positive coefficient which depends upon the parameters 
of the model. This relationship is consistent with $\epsilon$ having a power-law relation to 
the number of trajectories, $\epsilon\sim M^{-\gamma}$, but (as illustrated by figure \ref{fig: 7}) the 
approach to this limiting power-law can be so slow that the exponent cannot be seen in numerical 
simulations.

As well as having a power-law dependence, $\epsilon\sim M^{-\gamma}$, in the limit as $M\to \infty$, 
there is a transient behaviour at finite values of $M$. We showed that this transient 
behaviour can be described quite accurately by a rather brutal approximation of the equations
describing our model, replacing a random walk with a Wiener process. 

We argued that varying parameters of this model system can cause a transition, from a 
phase in which $\gamma$ is positive, to regions of parameter space in which it is zero. 
It is, however, difficult to observe a sharp phase transition upon varying parameters 
of the model, because the width of the transition region, where the limiting slope of the 
plot of $\langle \ln \epsilon \rangle$ versus $\ln M$ becomes established, increases 
as $\gamma\to 0$. Equations (\ref{eq: 3.3.1}), (\ref{eq: 3.3.2}) imply that the width of this transition region 
is $\ln M^\ast\sim D/v^2$, so that seeing the change of slope in the transition region 
required a very large number of trajectories, $M^\ast\sim \exp(2D/v^2)$.  

In the introduction we mentioned that the concept of multifractal measures appears 
as if it may be relevant to our investigation. The power-law relation $\epsilon\sim M^{-\gamma}$ is 
consistent with the \lq multifractal' model, in that it represents an exponent which characterises 
the dimension of the measure in the vicinity of a point. However, the exponent $\gamma$ is the same for 
almost all points in the phase space, rather than different values 
of $\gamma$ being realised on sets which have a fractal 
structure. Also, as illustrated in figure \ref{fig: 7}, the convergence 
towards a power-law as the number of trajectories increases can be so slow 
that it is not observable.
 
Figure \ref{fig: 1} showed voids in the distribution of a physically interesting invertible, non-autnomous 
chaotic system. It is interesting to consider how the approach used on our simplified model 
can be extended to understand the distribution of the distance $\varepsilon$ from a randomly
chosen point to the nearest one of $M$ trajectories in more general cases. 
As in the case of the simplified model that we have considered 
here, the simplest way to understand the statistics of $\varepsilon$ is to propagate the dynamics 
backwards in time. All of the pre-images of this $\varepsilon$ ball are also empty. 
In particular, the pre-image at time zero does not contain any of the initial random 
distribution of trajectories. Because the trajectories were assumed to be randomly scattered at 
time zero, the pre-image set at $t=0$ is very unlikely to have an area which exceeds $1/M$ by a large 
factor. 

Consider that form of the pre-images of a small ball of radius $\varepsilon$. The evolution 
of this set under backward time evolution is, at least initially, described by the 
linearisation of the flow. In many examples, including the case illustrated in figure \ref{fig: 1},  
the pre-image of a ball is initially transformed into an ellipse with one principal axis increasing 
and the other one decreasing, such that the area is contracting. 
Eventually, the linearisation approximation fails, when the size of the larger principal axis 
of the ellipse approaches unity. Upon further backward propagation, the pre-image set is a 
string-like object, which eventually becomes foliated so that it covers the whole of the 
phase space with uniform density. When this happens, the area remains 
approximately constant as we propagate backwards in time, because the dynamics 
preserves the total area. This picture is quite analogous to our treatment of the 
strudel model, but the machinery of the calculations will be more complex. 
We expect to explore the generalisation to more complex dynamical systems in a subsequent 
paper. 

\section*{Acknowledgements}
\label{sec: ack}

MW acknowledges hospitality of the Chan-Zuckerberg Biohub, and discussions with 
John Hannay about different approaches to the derivation of equation (\ref{eq: 3.3.3}). 
MP acknowledges financial support by the UK Engineering and 
Physical Sciences Research Council (EPSRC) through Grant No. EP/R041954/1.

\section*{Appendix: Derivation of expectation value of maximum of Wiener process}
\label{sec: A}

In \cite{Gra+15} there is an analysis of the solution of the advection diffusion equation, 
with drift velocity $v$ and diffusion coefficient $D$. It is shown that the flux onto an absorbing 
point at $\bar x$ from a source at $x=0$, $t=0$ is
\begin{equation}
\label{eq: A.1}
J(\bar x,t)=\frac{\bar x}{\sqrt{4\pi Dt^3}}\exp\left[-\frac{(\bar x-vt)^2}{4Dt}\right]
\ .
\end{equation}
The probability that a particle has a maximum excursion which is 
less that $\bar x$ before time $t$ is equal to the probability that it is 
not absorbed onto that surface, namely
\begin{equation}
\label{eq: A.2}
P(\bar x,t)=1-\int_0^t{\rm d}t'\ J(\bar x,t')
\ .
\end{equation}
The corresponding probability density for $\bar x$ is $p(\bar x,t)=\partial P/\partial \bar x$, 
so that the expectation value of $\bar x$ is
\begin{equation}
\label{eq: A.3}
\langle \bar x\rangle =-\int_0^t{\rm d}t'\int_0^\infty {\rm d}\bar x\ \bar x\frac{\partial J}{\partial \bar x}(\bar x,t')
=\int_0^t{\rm d}t'\int_0^\infty {\rm d}\bar x\ J(\bar x,t')
\ .
\end{equation}
That is, defining $Z=v\sqrt{t'/2D}$ and $Y=v\sqrt{t/2D}$, 
\[
\langle\bar{x}\rangle=\frac{1}{\sqrt{4\pi D}}\int_{0}^{t}{\rm d}t'\ \frac{1}{t'^{3/2}}\int_{0}^{\infty}{\rm d}
\bar{x}\ \bar{x}\exp\left[-\frac{(\bar{x}-vt')^{2}}{4Dt'}\right]
\]
\[
=\frac{1}{\sqrt{2\pi}}\int_{0}^{t}{\rm d}t'\ \frac{1}{t'}\int_{-v\sqrt{t'/2D}}^{\infty}{\rm d}\omega\ 
\left[\sqrt{2Dt'}\omega+vt'\right]\exp\left(-\frac{\omega^{2}}{2}\right)
\]
\[
=\sqrt{\frac{D}{\pi}}\int_{0}^{t}\frac{{\rm d}t'}{\sqrt{t'}}\int_{-Z}^{\infty}{\rm d}\omega\ \omega
\exp\left(-\frac{\omega^{2}}{2}\right)+\frac{v}{\sqrt{2\pi}}\int_{0}^{t}{\rm d}t'\int_{-Z}^{\infty}{\rm d}\omega\ 
\exp\left(-\frac{\omega^{2}}{2}\right)
\]
\[
=\sqrt{\frac{D}{\pi}}\int_{0}^{t}\frac{{\rm d}t'}{\sqrt{t'}}\exp\left(-\frac{Z^{2}}{2}\right)+v\int_{0}^{t}{\rm d}t'\ \Phi(Z)
\]
\[
=\frac{4D}{v}\frac{1}{\sqrt{2\pi}}\int_{0}^{Y}{\rm d}Z\ \exp\left(-\frac{Z^{2}}{2}\right)
+\frac{4D}{v}\int_{0}^{Y}{\rm d}Z\ Z\,\Phi(Z)
\]
\[
=\frac{2}{Y}\sqrt{2Dt}\left[\Phi(Y)-\frac{1}{2}+\frac{Y^{2}}{2}\Phi(Y)
-\frac{1}{2\sqrt{2\pi}}\int_{0}^{Y}{\rm d}Z\ Z^{2}\exp\left(-\frac{Z^{2}}{2}\right)\right]
\]
\[
=\sqrt{2Dt}\left[\frac{2}{Y}\Phi(Y)-\frac{1}{Y}+Y\Phi(Y)
-\frac{1}{Y\sqrt{2\pi}}(-Y\exp(-\frac{Y^{2}}{2})+\sqrt{\frac{\pi}{2}}(2\Phi(Y)-1))\right]
\]
\[
=\sqrt{2Dt}\left[\Phi(Y)(Y+\frac{1}{Y})+\Phi'(Y)-\frac{1}{2Y}\right]
\]
\begin{equation}
	=\sqrt{2Dt}\, F(Y)
\end{equation}
where $F(Y)$ is the function specified in equation (\ref{eq: 3.3.3}).
There are other sources which could be used to obtain (\ref{eq: 3.3.1}) and (\ref{eq: 3.3.3}), 
for example a book by Borodin and Salminen (\cite{Bor+02}, see Part II, ch.2, eq. (1.1.4), p.250), 
although there is an error in the published formula.

\end{document}